\documentclass[twoside]{dis09}
\usepackage[latin1]{inputenc}
\usepackage[dvips]{graphicx,epsfig,color}
\usepackage{wrapfig,rotating}
\usepackage{amssymb,amsmath,array}

\begin{document}
\title{Structure Functions and Low-{\boldmath $x$}\\ 
   Working Group Summary}

\author{Burkard Reisert$^1$, Agustin Sabio Vera$^2$ and Zhiqing Zhang$^3$
%
%
\vspace{.3cm}\\
%
1- Max-Planck-Institut f\"ur Physik \\
F\"ohringer Ring 6, 80805 M\"unchen - Germany
%
\vspace{.1cm}\\
2 - Instituto de F{\' i}sica Te{\' o}rica UAM/CSIC,\\ 
Universidad Aut{\' o}noma de Madrid, E-28049 Madrid - Spain
%
\vspace{.1cm}\\
3- Laboratoire de l'Accélérateur Linéaire \\
Université Paris-Sud 11 et IN2P3/CNRS, Orsay - France
}

\maketitle

\vspace{-11pt}
\begin{abstract}
A summary of recent results reported in the Structure Functions and Low-$x$ working group at the DIS 2009 Workshop is given.
%
\end{abstract}

\vspace{-6pt}
\section{Introduction}
\vspace{-10pt}
Nucleon structure functions and their scale variations are closely related 
to the origins of Quantum Chromodynamics (QCD) as a gauge theory of the 
strong interaction. Precision data from HERA and the Tevatron lay the 
foundation of a quantitative understanding of the nucleon's structure 
in terms of parton distribution functions and their uncertainties.

\vspace{-5pt}
\section{Deep Inelastic Scattering}
\vspace{-10pt}
Since the completion of the HERA program at DESY considerable progress has 
been made to perfect the measurement of the inclusive electron/positron proton
cross sections. An impressive amount of published and new preliminary neutral 
and charged current inclusive cross section and structure function data
covering the entire HERA kinematic regime became available over the year passed
since the DIS meeting in 2008.

\vspace{-5pt}
\subsection{Precision Measurements at Low and Medium-{\boldmath $Q^2$}}
\vspace{-5pt}
A new measurement~\cite{H1:medQ2new} of the inclusive positron proton 
cross section using approximately 22\,pb$^{-1}$ of data recorded by the H1 
detector in 2000 was presented by Kretzschmar. The new measurement covers 
a range of medium four momentum transfer squared, $12 < Q^2 < 150$\,GeV$^2$ and 
inelasticity $y<0.6$. For the reconstruction of the kinematic 
variables, two methods are employed, one relying solely on the measurement 
of the scattered electron, the other also including information about the
hadronic final state. For the final measurement, the reconstruction 
method which yields the smallest systematic uncertainty is used. 

Petrukhin reported an extention of the H1 inclusive cross section 
measurements ~\cite{H1:lowQ2new}
towards low-$Q^2$, $0.2 < Q^2 < 12 $GeV$^2$. This measurement is based 
on a dedicated shifted vertex run which improved the
detector acceptance for low-$Q^2$ and a minimum bias 
run with open triggers for low-$Q^2$ inclusive data. 
Both runs were done in 1999 and 2000 during the HERA-I period.

The new H1 measurements at medium and low-$Q^2$, 
are combined with the published H1 results~\cite{H1:lowQ2} using 
an averaging procedure which takes into account correlated systematic uncertainties. 
The combination of the measurements not only yields smaller statistical errors but also
a better understanding of the systematic uncertainties. The combination and 
reanalysis of the old data revealed a small, $Q^2$-dependent bias in the old data
of up to 2.5\%. After the combination an unprecedented accuracy of 1.3\% to 2\% 
is achieved. 

\begin{figure}[htb]
  \centerline{
\vspace{-15pt}
    \includegraphics[width=0.5\columnwidth,height=5.5cm]{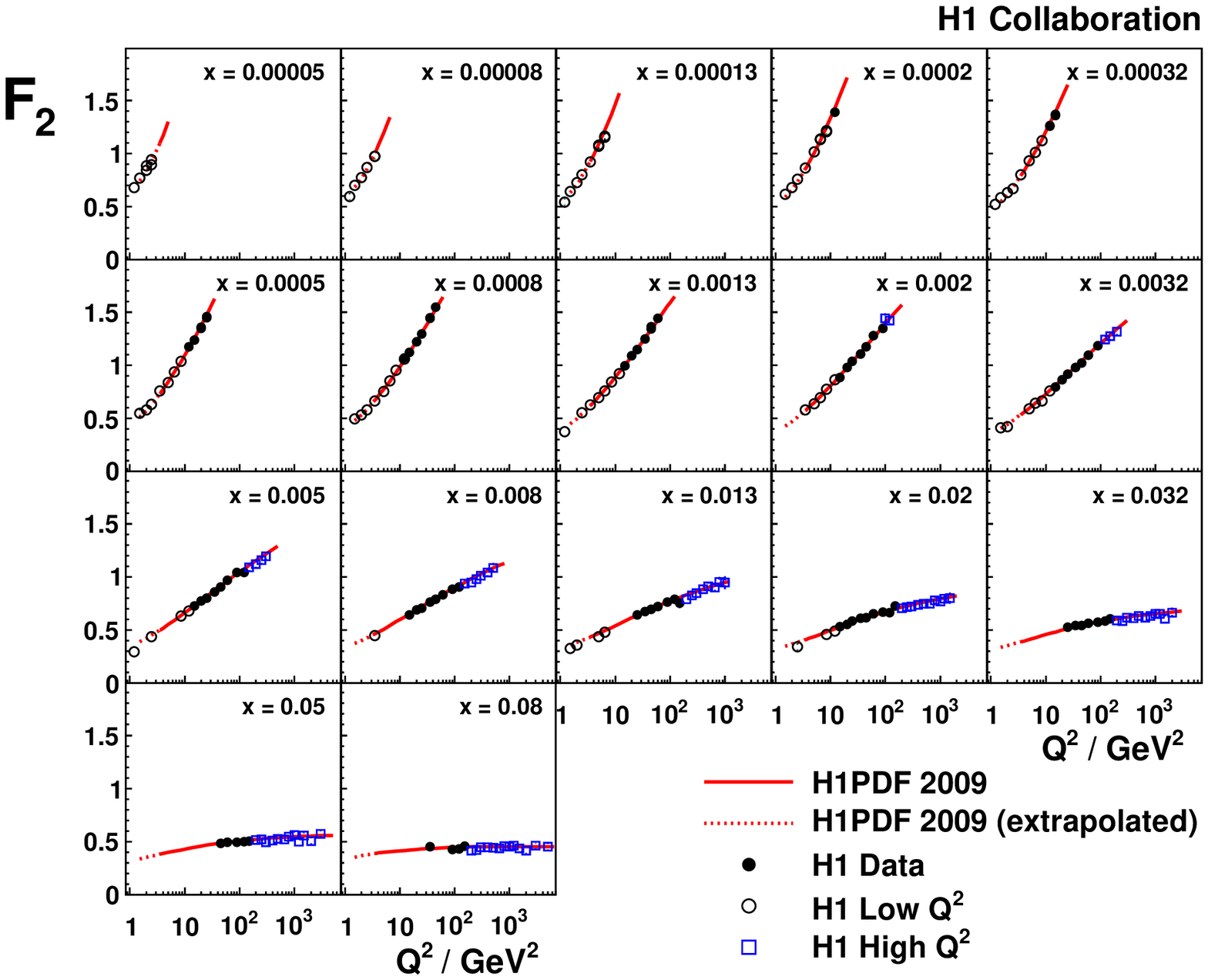}
    \includegraphics[width=0.45\columnwidth,height=5.5cm]{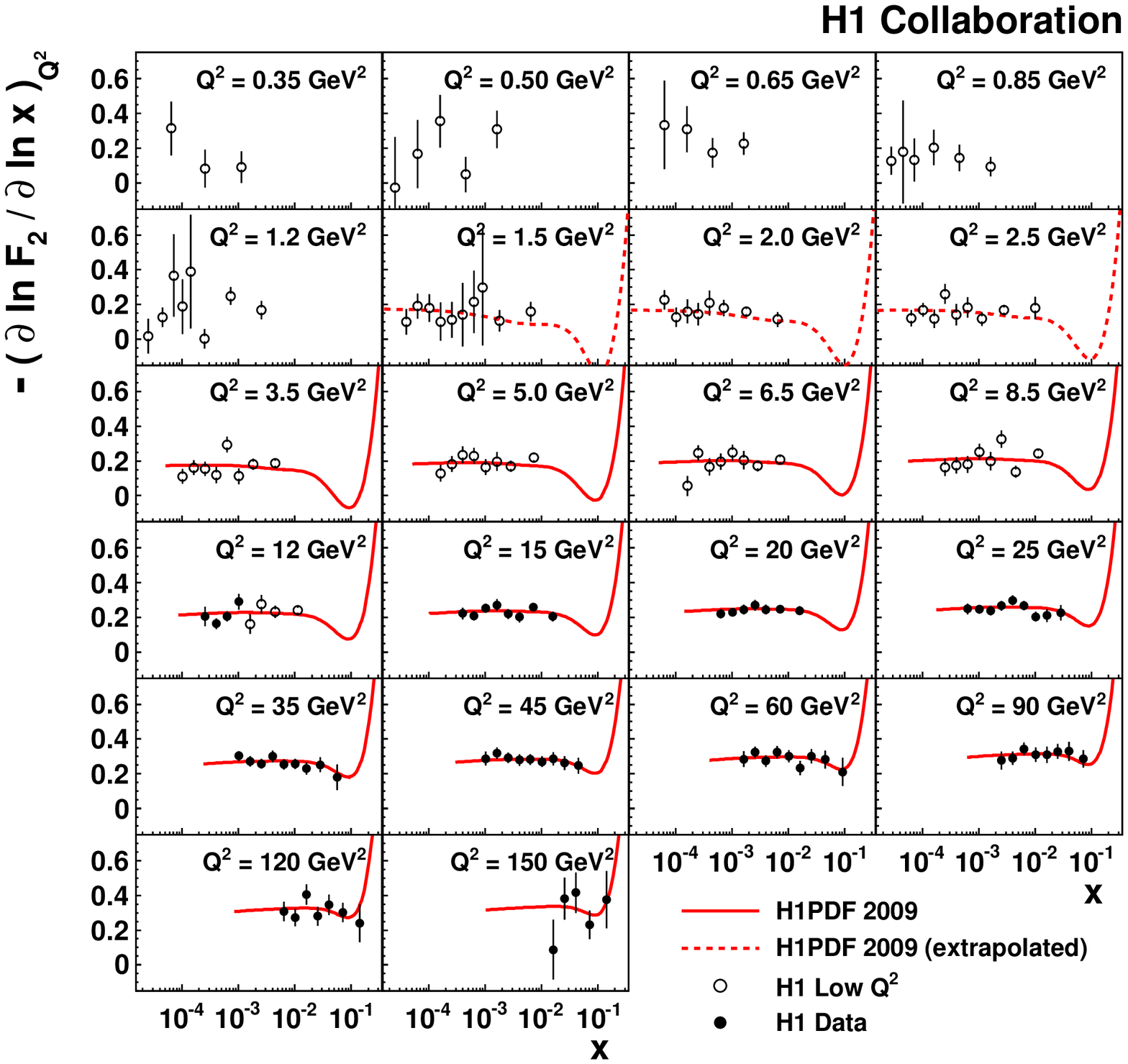}
}
\caption{Measurement of the proton structure function $F_2$ as a function of $Q^2$ for 
fixed values of $x$ (left) and of the derivative $\partial \ln F_2/\partial\ln x|_{Q^2}$
as a function of $x$.}
\label{Fig:H1F2}
\vspace{-12pt}
\end{figure}

The accurate measurements of the cross section allow for an extraction of the
proton structure function $F_2$ and its derivatives, as shown in Fig.~\ref{Fig:H1F2}. 
In the perturbative regime
($Q^2 \geq {\cal O}(1\,{\rm GeV}^2)$) the measurements are well described by NLO predictions.
At lower $Q^2$, i.e. in the transition region towards non-perturbative QCD, 
phenomenological models (e.g. the colour dipole model\cite{Dipole:GBW,Dipole:IIM} or
a fractal fit~\cite{Fractal} based on the concept of self similarity) are found to give a 
decent parameterisation of the measurement. At low Bjorken $x$, $F_2$ can be 
parameterised as $F_2 = c(Q^2) x^{-\lambda(Q^2)}$ allowing for an extrapolation
of $F_2$ towards even lower $x$ i.e. higher $y$ enabling an indirect determination
of the structure function ratio $R=F_L/(F_2-F_L)$. The value of averaged $R$ 
for $Q^2 < 12$\,GeV$^2$ extracted in this model dependent way is found to be 
consistent with $R=0.5$, which is twice higher than that obtained from direct measurements
of $F_L$, see the following section.

\subsection{Direct Measurement of the Longitudinal Structure Function $F_L$}
\begin{wrapfigure}{r}{0.425\columnwidth}
\vspace{-15pt}
\centerline{\\[-5mm]\includegraphics[width=0.4\columnwidth]{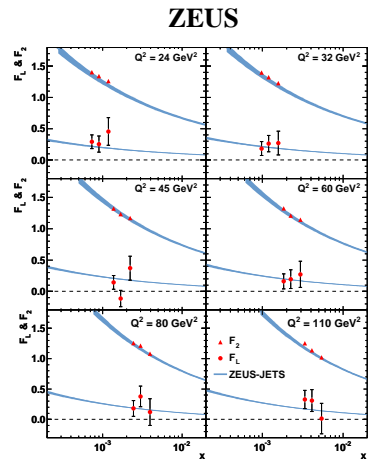}}
\caption{ZEUS $F_L$ and $F_2$ as a function of $x$ at fixed $Q^2$.}
\label{Fig:ZEUS:FL} 
\vspace{-20pt}
\end{wrapfigure}
The reduced NC $ep$ inclusive cross section is expressed in terms of structure functions by 
$\sigma_r = F_2(x,Q^2) - y^2/Y_+ F_L(x,Q^2)$ with $Y_+=1+(1-y)^2$.
The structure functions $F_2$ and $F_L$ can be extracted from the measurement of $\sigma_r$ at fixed $x$ and $Q^2$ 
but varying $y$. The kinematic variables $x$, $Q^2$ and $y$ are related by $y = \frac{Q^2}{x\cdot s}$ 
where $s$ is the electron proton centre-of-mass energy squared. Therefore varying $y$ -- while 
keeping $x$ and $Q^2$ constant -- requires to vary $s$. At HERA the variation of $s$ was achieved by 
lowering the proton energy from 920\,GeV to 460\,GeV in the so-called low energy run, LER, and to 575\,GeV for the medium energy run, MER. 
During a dedicated running period at the end of the HERA program both colliding beam experiments, H1 and ZEUS
record approximately 14\,pb$^{-1}$ of LER and 7\,pb$^{-1}$ of MER data. 

Both H1 and ZEUS recently published first results of the longitudinal structure function $F_L$ at medium $Q^2$\cite{H1:FL,ZEUS:FL}.
The ZEUS analysis, presented by Grebenyuk, combines the LER and MER with cross sections measured with 
approximately 44\,pb$^{-1}$ at the nominal centre-of-mass energy. The latter constitutes the most precise 
cross section measurement of ZEUS in the kinematic region covered. The extracted values of $F_L$ and $F_2$ 
displayed in Fig.~\ref{Fig:ZEUS:FL} are well described by the NLO prediction based on the ZEUS-JETS fit.  
For the $Q^2$ range covered, 24 to 110\,GeV$^2$, ZEUS quotes a value 
of $R = 0.18^{+0.07}_{-0.05}$. 

\begin{wrapfigure}{r}{0.475\columnwidth}
\vspace{-10pt}
\centerline{\\[-3.5mm]\includegraphics[width=0.475\columnwidth]{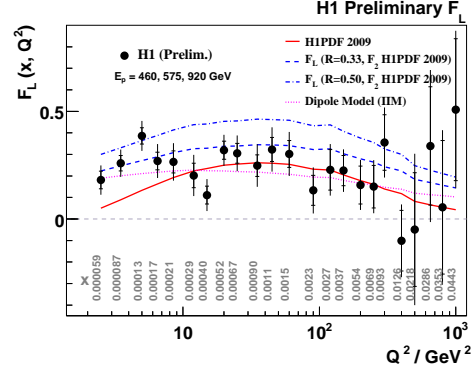}}

\caption{H1 average $F_L$ as a function of $Q^2$.}
\label{Fig:H1:FL} 
\vspace{-5pt}
\end{wrapfigure}
The H1 analysis, presented by Glazov, profited from the continued efforts to improve the 
backward calorimetry and tracking systems as well as from various trigger upgrades to enhance the
trigger efficiency of low energy electrons. This enables H1 to measure the scattered 
electron at an energy as low 
as 3\,GeV, while using the charge measurement of the track associated with the electron candidate
to control the background. H1 have extended their published measurement 
towards high-$Q^2$, up to 800\,GeV$^2$, 
and low-$Q^2$, down to 2.5\,GeV$^2$, and obtain values of $F_L$ at up to 6 different values of $x$ 
at a given $Q^2$.  The $F_L$ averaged over $x$ as a function of $Q^2$ is shown in Fig.~\ref{Fig:H1:FL}. For 
$Q^2 > 10$\,GeV$^2$, the measurement is well described by NLO predictions, whereas at lower $Q^2$ the
perturbative QCD calculation underestimates the measurement. Dipole models are found to describe 
the data well. For the entire $Q^2$-range the measurement is consistent with a value of $R=0.25$.   

\subsection{Neutral and Charged Current Cross Sections at High-{\boldmath $Q^2$}}
\begin{wrapfigure}{r}{0.475\columnwidth}
\centerline{\includegraphics[width=0.475\columnwidth]{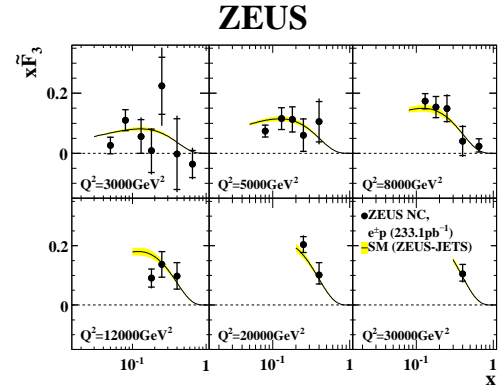}}
\caption{The structure function $xF_3$ as a function of $x$ at fixed $Q^2$.}
\label{Fig:ZEUS:xF3} 
\end{wrapfigure}
New measurements by ZEUS of neutral (NC) and charged current (CC) cross sections at high-$Q^2$ were
presented by Cooper-Sarkar. ZEUS recently published the measurement of NC~\cite{ZEUS:e-p:NC} and CC~\cite{ZEUS:e-p:CC} $e^-p$ cross sections
based on 169\,pb$^{-1}$ and 175\,pb$^{-1}$. This new data samples are  approximately 10 times larger 
than those of the previous publication from the HERA-I period. This enables a much improved measurement of
the structure function $xF_3$, see Fig.~\ref{Fig:ZEUS:xF3}. In addition preliminary measurements
of the NC and CC $e^+p$ cross sections were presented. Since the HERA upgrade the machine 
operated with polarised electron/positron beams. For both lepton beam charges, measurements with 
positive and negative polarisation are now available, and effects of the polarisation 
have been clearly established both for CC and NC processes, giving striking confirmation of the
Standard Model prediction for the chiral structure of the weak interactions.

\subsection{Combined Cross Section of HERA-I } 
Tassi reported on the update of the combination of cross section measurements from 
H1 and ZEUS applying the averaging procedure presented at 
DIS\,2008~\cite{H1:ZEUS:AVG}. In addition to the data already incorporated 
last year, the updated HERA combined cross section 
now also include the H1 low and medium $Q^2$ cross sections (see above) 
and the low $Q^2$ data recorded with the small angle ZEUS beam-pipe 
calorimeter (BPC~\cite{ZEUS:BPC}) and beam-pipe tracker (BPT~\cite{ZEUS:BPT}) as well as data 
from ZEUS recorded 
during a dedicated data taking period with shifted nominal vertex position~\cite{ZEUS:SVX} to increase the acceptance of small polar angles of the scattered electron. 
The ZEUS and H1 data are shifted to a common $x$-$Q^2$-grid. Relying on the 
trivial assumption that both experiments are bound to measure
the same cross section at the same $x$ and $Q^2$, 
the averaging procedure not only reduces the statistical errors but also 
reduces the systematic uncertainty of the combined cross section. 
The global fitting groups expressed their great interest that the averaged 
HERA-I cross sections, which now comprise the complete set of the HERA-I inclusive 
measurements, are made available in order to be incorporated in the
global fits.

\subsection{Structure Function Measurements at HERMES}
Gabbert reported on new measurements of the proton and 
deuteron structure functions $F_2^p$ and $F_2^d$ by the HERMES collaboration, 
which uses the HERA electron/positron beam in fixed target mode.  
The results are based on a data sample which is 10 (5) times larger 
than that of the 
\begin{figure}[htb]
\centerline{
  \includegraphics[width=0.45\columnwidth]{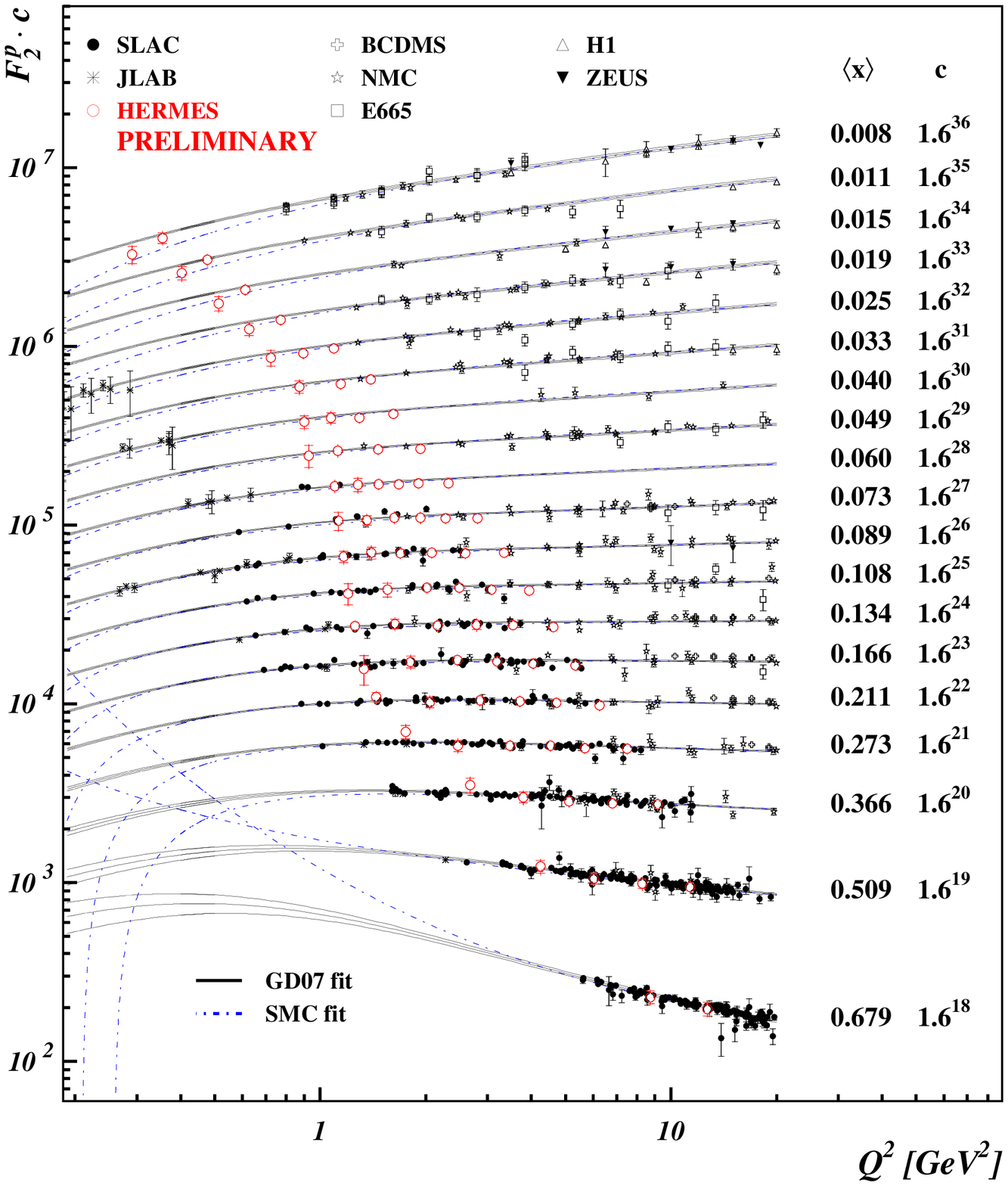}
  \hspace{5mm}
  \includegraphics[width=0.45\columnwidth]{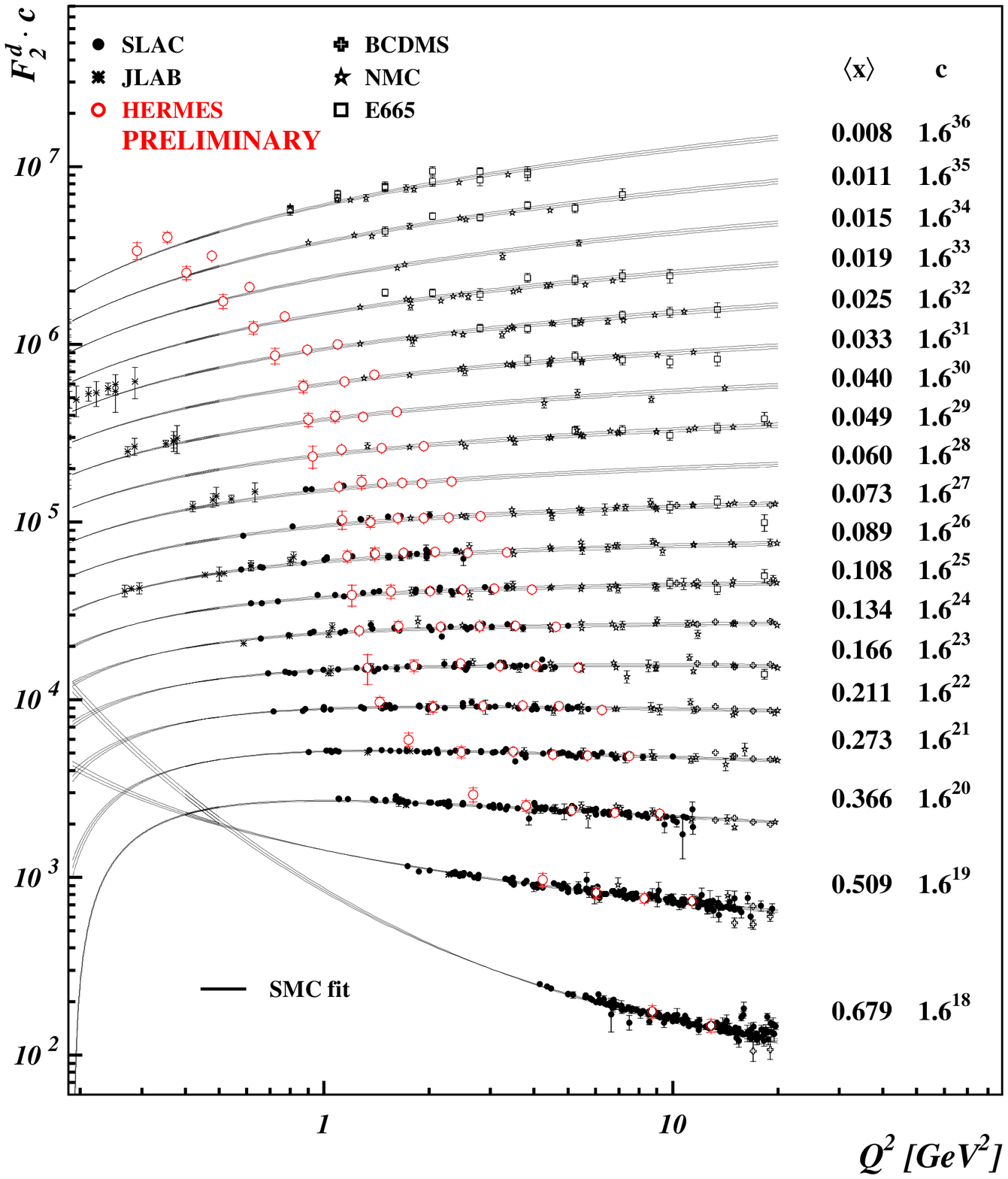}
}
\caption{Measurement of the proton structure function $F_2^p$ (left) and
deuteron structure function $F_2^d$ (right) by the HERMES collaboration.}\label{Fig:HERMES}
\end{figure}
proton (deuteron) structure function measurements by NMC with which this 
new measurements overlap. The measurements are found to be consistent 
with previous measurements. Towards low $x$ and $Q^2$ the HERMES data 
explore a new kinematic region hitherto uncovered by experimental data.
A new measurement of the cross section 
ratio $\sigma^d/\sigma^p$ was obtained.
It will be interesting to see the impact of this new data in future global fits,
as well as its impact on extractions of the Gottfried Integral and the  
valance quark ratio ($d_v/u_v$) at high $x$.

\vspace{5pt}
\section{Proton Anti-Proton Collisions}
\vspace{1pt}
Results from the Tevatron experiments, CDF and D\O, were presented 
by McNulty and Fox. The Tevatron $p\bar{p}$ data provide important 
constraints for $d_v$ and $u_v$ valence quarks via the measurement 
of W and Z production and of the gluon distribution at high $x$ via 
the measurement of inclusive jet cross sections. Recently both experiments 
invested quite some effort to extend their measurements to larger 
rapidities $\eta$ thus simultaneously probing the high- and low-$x$ domain.
The extension to high rapidities ($\eta \sim 2$) at the Tevatron helps 
constraining standard model processes at the LHC at central rapidities ($\eta = 0$). 

\vspace{1pt}
\subsection{Inclusive Jet Cross Sections}
\vspace{1pt}
\begin{wrapfigure}{r}{0.425\columnwidth}
\centerline{\includegraphics[width=0.425\columnwidth,height=4.5cm]{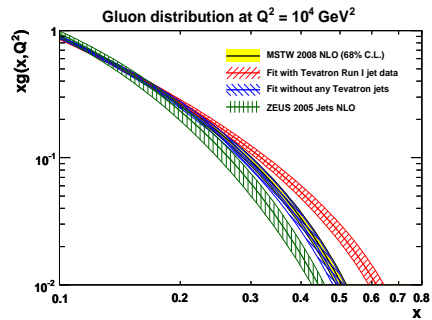}}
\caption{Comparison of high-$x$ gluons obtained from PDF fits with 
varying jet input data~\cite{MSTW08}.}\label{Fig:GluonJet} 
\end{wrapfigure}
Both experiments presented  measurements of the double differential 
cross sections $d^2\sigma/dp_{T}dy$ using 0.7\,fb$^{-1}$ for the D\O\ and 
1.13\,fb$^{-1}$ for the CDF analysis. The measurements cover a $p_T$-range
from 50 to 600\,GeV and extend to a rapidity of 2.1 (CDF) and 2.4 (D\O).
Within the systematics of the measurements the results of CDF~\cite{CDF:Jets:Kt,CDF:Jets:Mid} and D\O~\cite{D0:Jets} are in 
mutual agreement and are also well described by NLO QCD 
calculations.
The impact of jet measurements on the gluon at high-$x$ has been studied 
in dedicated PDF fits 
by varying the jet datasets included in the fit, yielding
variations of the gluon central value larger than the uncertainty bands 
obtained in the individual fits, examples of the resulting gluon are shown in Fig.~\ref{Fig:GluonJet}. 
A detailed understanding of the
gluon at high-$x$ requires further dedicated study.  

\vspace{1pt}
\subsection{$W^\pm$ Charge Asymmetry}
\begin{figure}[htb]
\centerline{
 \includegraphics[width=0.475\columnwidth,height=5.25cm]{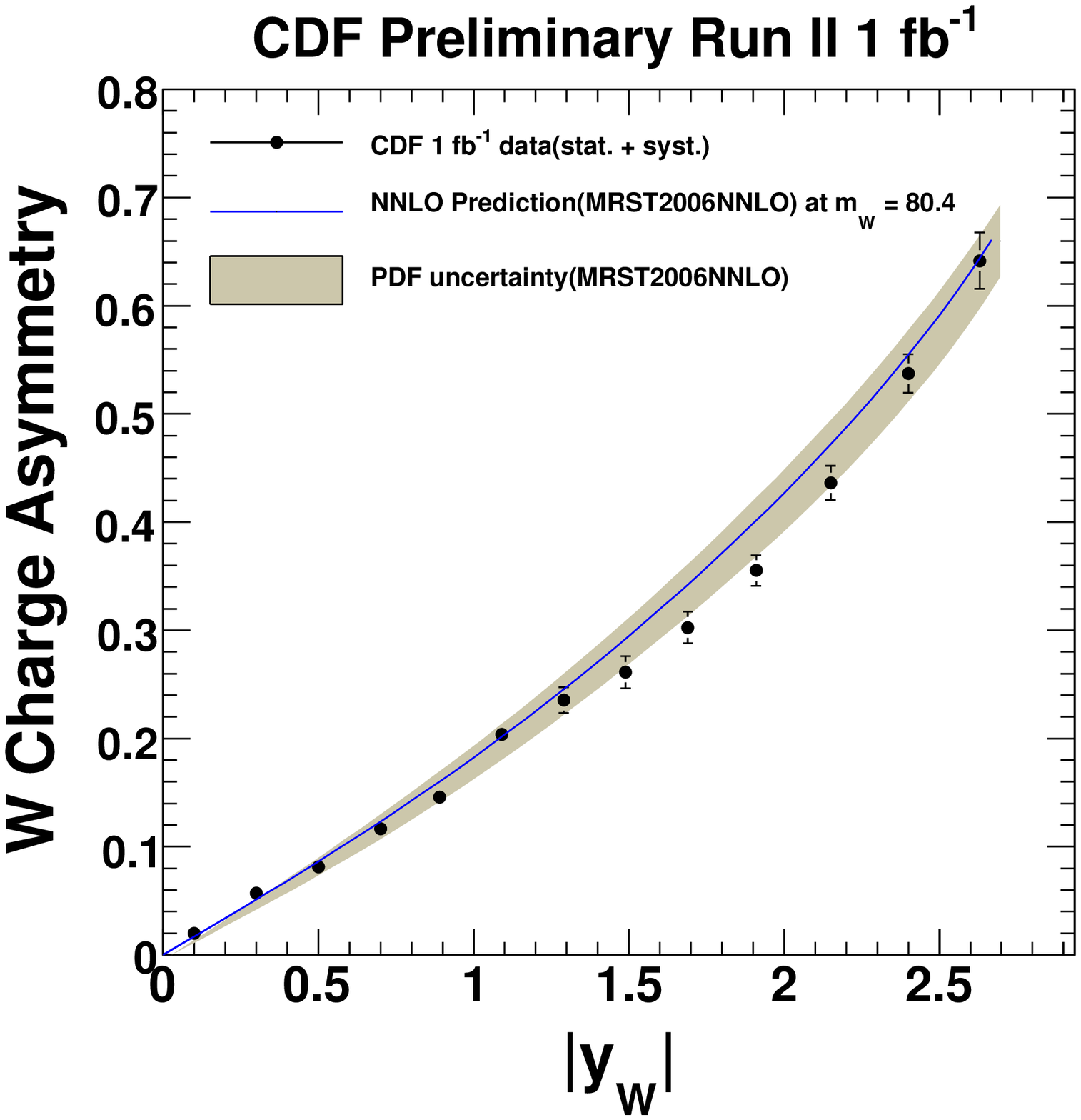}
 \includegraphics[width=0.475\columnwidth]{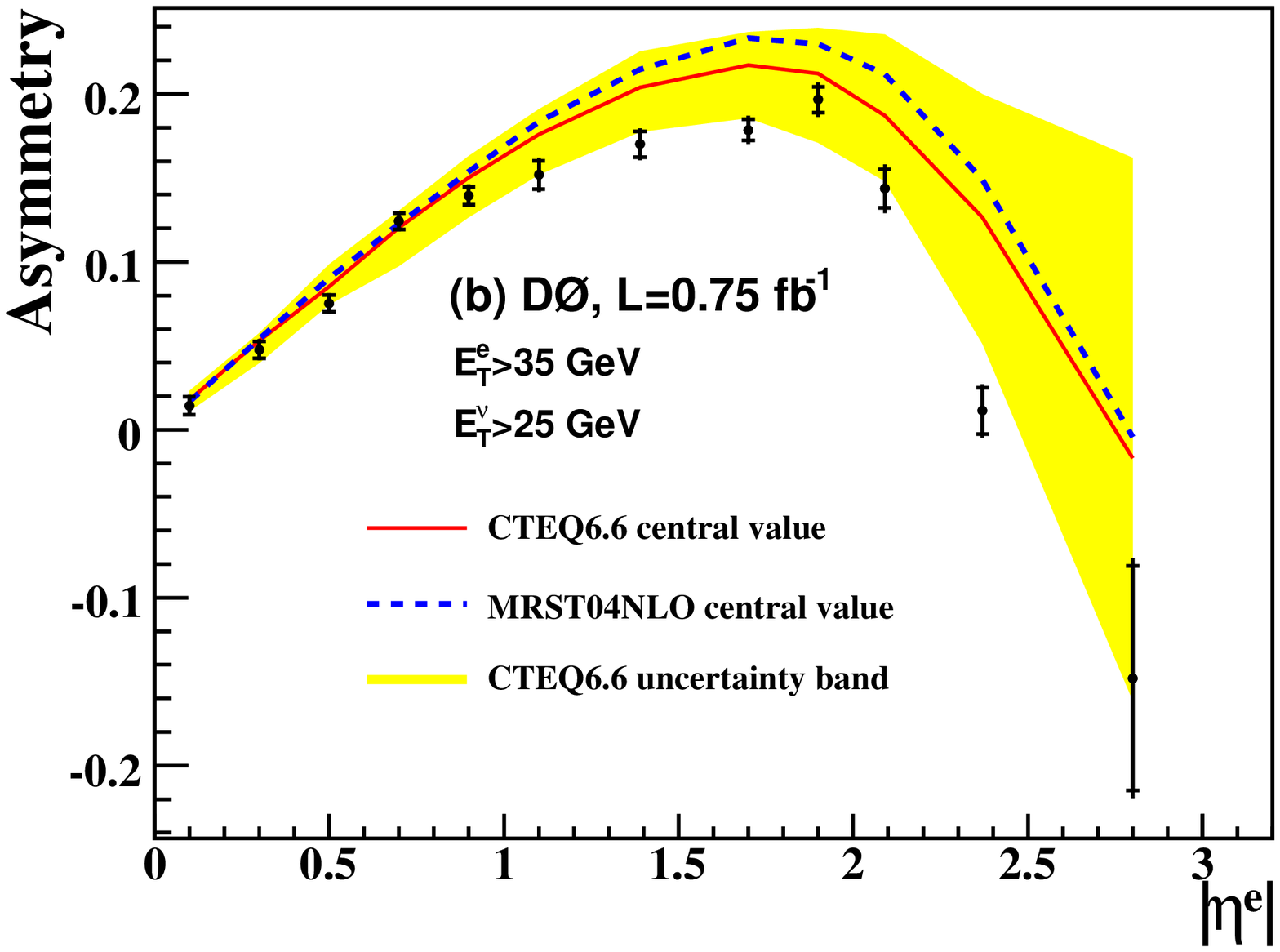}
}
\caption{Measurements of the $W^\pm$ charge asymmetry as a function
of rapidity of the $W$ boson (CDF, left) and of $W$ decay 
lepton (right, D\O).}\label{Fig:TevW}
\end{figure}

\vspace{1pt}
At the Tevatron $W$ bosons are produced mainly from a quark of the
proton and an anti-quark of the anti-proton. On average the $u$-quark
momentum fraction is larger than the $d$-quark. Therefore the $W^+$ 
($W^-$) is preferentially boosted along the (anti-)proton direction.
Both experiments presented measurements of the $W^{\pm}$ 
charge asymmetry measurement based on their Tevatron Run II data
using the standard approach which relies on the rapidity of the
W decay lepton. CDF also employs a method which directly reconstructs 
the $W^\pm$ rapidity  with up to two-fold ambiguity based on the 
W mass as a constraint~\cite{TeV:W:Tech}. Results from both experiments are shown 
in Fig.~\ref{Fig:TevW}. 
The uncertainties of the measurements using 1\,fb$^{-1}$ (CDF) and 
0.76\,fb$^{-1}$ (D\O) are smaller than the uncertainty attributed to the PDFs 
in the prediction, indicating the constraining power of this type of
measurement.

\subsection{$Z$ Boson Production}

\begin{wrapfigure}{r}{0.5\columnwidth}
\centerline{\includegraphics[clip=,width=0.5\columnwidth,height=5cm]{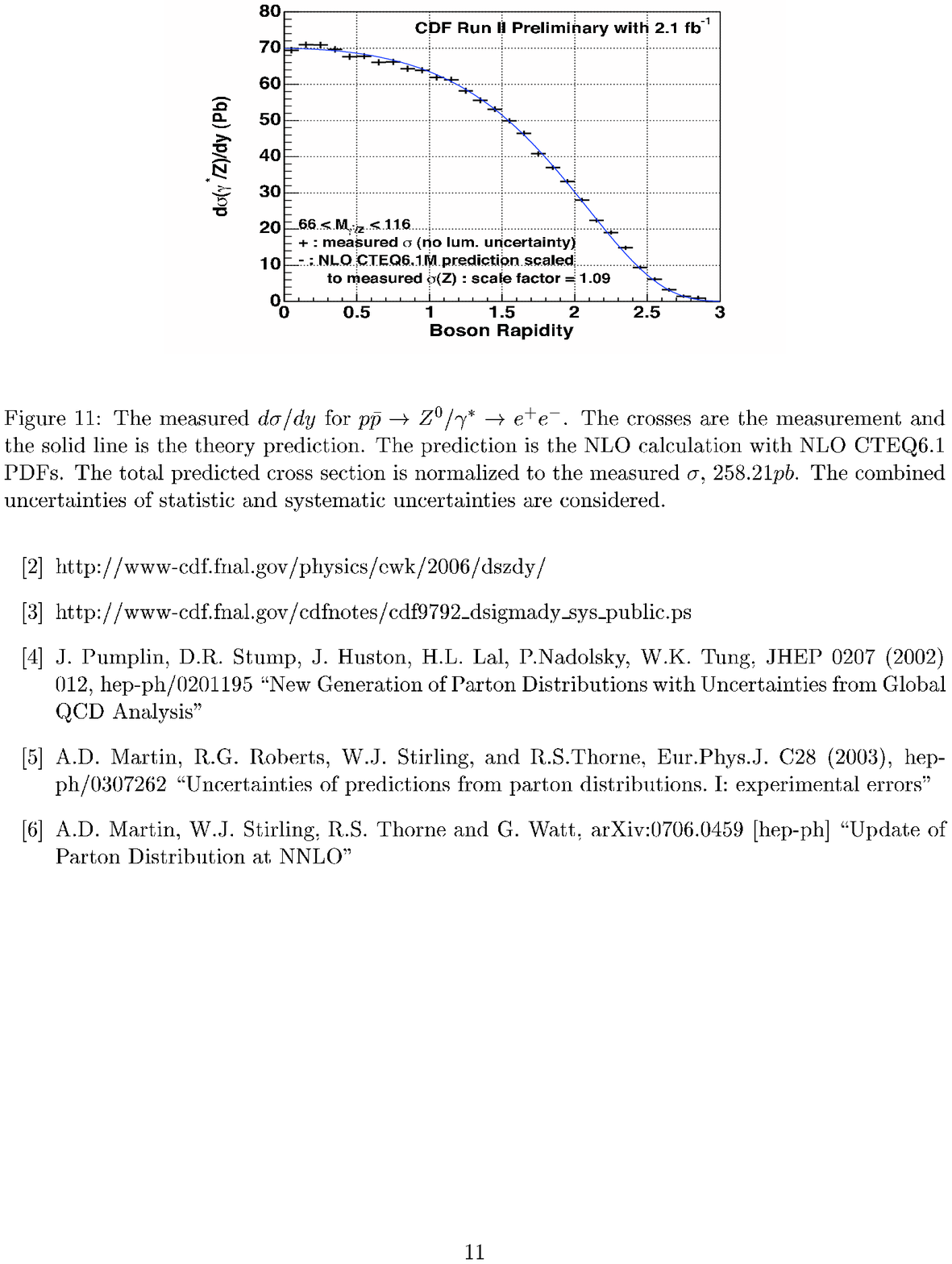}}
\caption{Rapidity distribution of $Z$ bosons measured by CDF 
compared to perturbative QCD predictions at NLO.}\label{Fig:TevZ} 
\end{wrapfigure}
At the Tevatron $Z$ bosons are produced via quark anti-quark annihilation.
The $Z$ boson rapidity is directly related to the momentum fractions of the
incoming partons of the hard interaction process.
CDF presented measurement of the $\gamma^*/Z$ production cross section, which 
extends to a boson rapidity of $y=3$. This new measurement, shown in Fig.~\ref{Fig:TevZ}, 
is based on 2.1\,fb$^{-1}$. D\O\ presented measurements of the $Z$ production cross section 
as a function of the transverse momentum of the $Z$-boson. D\O\ also presented a measurement of the
Z forward backward asymmetry, $A_{FB}$ based on 1.1\,fb$^{-1}$. At present the precision is 
still statistically limited, however in future, with 6 to 8\,fb$^{-1}$ anticipated for the 
Tevatron Run II, the measurement of $A_{FB}$ is expected to provide further constraints
on PDFs.   

\newpage  
\section{Extraction of Parton Densities}
Progress in obtaining an improved quantitative 
understanding of parton distribution functions
and their uncertainties was reported at the meeting. 

\subsection{Extractions of PDFs at HERA}
In both collaborations, H1 and ZEUS, the extraction of PDFs is an very 
active field of investigation. Naturally activities in both collaborations
are focusing on the HERA data. 

Cooper-Sakar presented an preliminary PDF analysis 
in the spirit of the ZEUS-JETS fit~\cite{ZEUSJETS} which incorporates
the new ZEUS measurements of NC and CC cross sections at high $Q^2$ 
as well as the high $y$ cross sections made available together with the
measurement of $F_L$, see above. While the high-$Q^2$ data sets reduce the
uncertainty of the valence quarks at high $x$, the medium-$Q^2$ high-$y$ 
data impact the gluon and sea quark distributions.

A new PDF fit to H1 data alone, labelled H1PDF2009~\cite{H1:medQ2new}, was presented by Kretzschmar. 
Radescu discussed the update of the PDF fit to the HERA combined cross sections, 
labelled HERAPDF0.2. In both studies incorporating the new more precise measurements 
reduces the experimental uncertainties of the extracted PDFs to a level that
model and parameterisation uncertainties become more important. In addition 
to the model uncertainties already considered in the earlier studies
 (H1PDF2000~\cite{H1PDF2000} and HERAPDF0.1~\cite{HERAPDF01}) the 
parameterisation uncertainty is estimated by 
calculating the envelope of PDF error bands obtained from fits of compatible quality 
with additional terms in the PDF parameterisations.

\subsection{Global PDF Fits}

In the context of best global fits Thorne reported on the recently published 
MSTW08 NLO DGLAP global fit~\cite{MSTW08}.
The main new ingredients to the fit are NuTeV and CCFR dimuon data, 
constraining the strange sea, inclusive jet data from HERA and the Tevatron, 
lepton asymmetry data and Z-boson rapidity measurements from CDF and D\O\ and 
all at the time of publication available charm structure function data. 
Considerable work went into the procedure to evaluate the PDF uncertainties
by applying a more sophisticated tolerance criterion to deal with tensions
between the input datasets.

Recent work of the CTEQ collaboration towards a new set of PDFs, CT09~\cite{CT09}, 
was reviewed by Nadolsky and Lai. The new analysis incorporates a variety 
of measurements from Tevatron with sensitivity to PDF, see above. The impact 
of the Tevatron Run-II jet data on the gluon distribution was studied 
in great detail including an critical review of jet cross section predictions 
which were found to be satisfactory (although not perfect)  at 
next to leading order. The compatibility of various data sets and 
sophisticated procedures to evaluate the PDF uncertainties are an 
active field of investigation.

Ubiali reported on the progress of the NNPDF collaboration towards a global PDF analysis~\cite{NNPDF}. 
Instead of parameterising the fitted PDFs with an arbitrarily chosen functional 
form this analysis employs neural networks, trained on Monte Carlo generated replica 
of the input data, to obtain a set of PDFs with an faithful estimate of the 
uncertainties~\cite{NNPDF:method}. A first set of PDFs from a comprehensive analysis of DIS data 
is now available, labelled NNPDF1.0. Work on including
data from hadron hadron collisions is progressing, detailed comparisons 
to the H1 and MSTW fit look very promising. A parton set from the NNPDF collaboration, 
which can compete with the global analyses by CTEQ and MSTW, appears to be 
feasible in the near future.

\subsection{Dedicated PDF Studies}
\begin{wrapfigure}{r}{0.45\columnwidth}
\vspace{-20pt}
\centerline{\includegraphics[clip=,width=0.45\columnwidth,height=8cm]{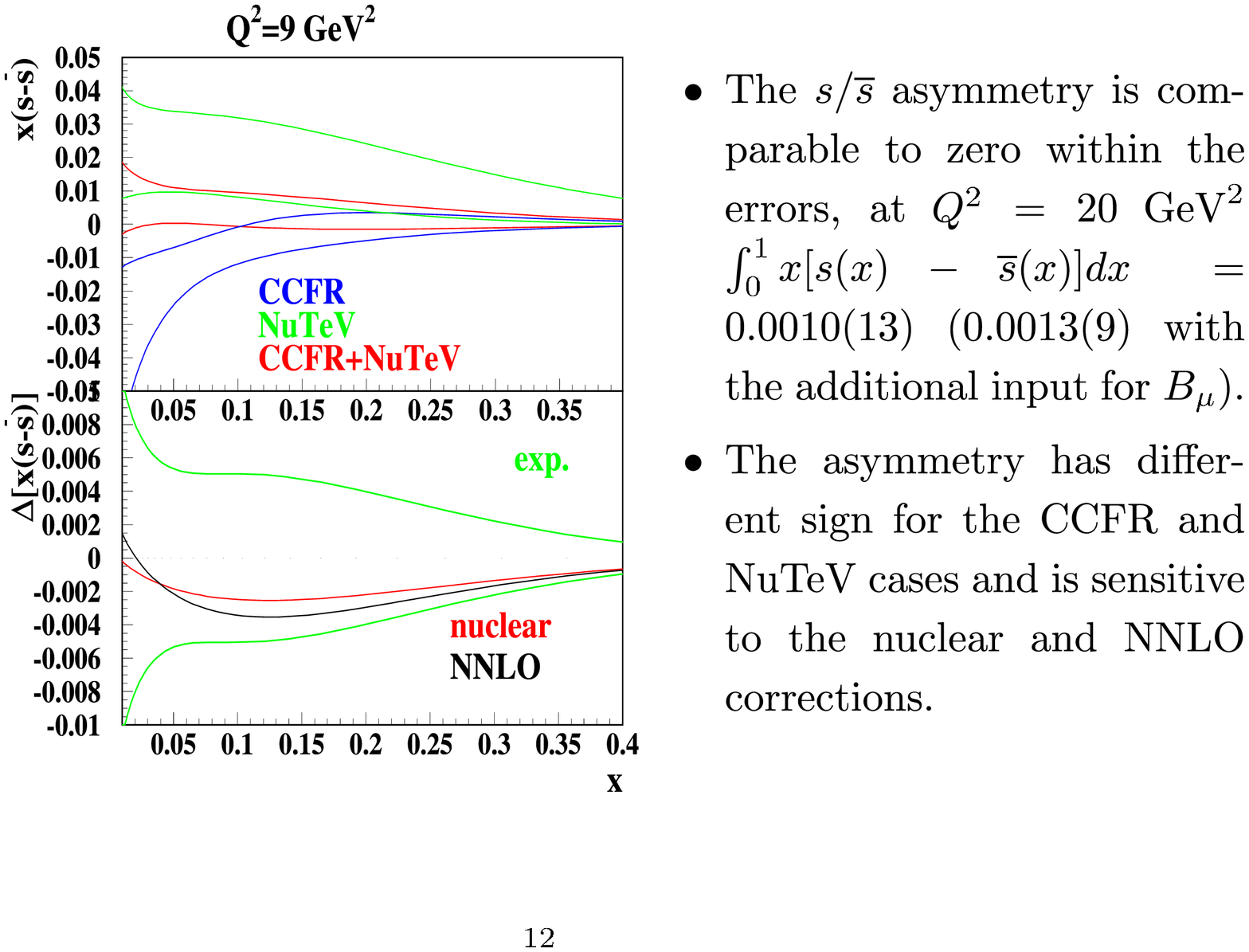}}
\caption{The strange sea asymmetry (top and its uncertainty (bottom) in various 
fit scenarios.}
\label{Fig:ssbar} 
\vspace{-20pt}
\end{wrapfigure}
Alekhin~\cite{Strange:Alekhin} and Rojo~\cite{Strange:Rojo} presented detailed studies of the strange sea. These studies
are motivated by the necessity of a flavour decomposition of the sea quarks
to achieve a faithful estimate of the sea quark uncertainties. In addition 
an asymmetric strange sea might well explain the NuTeV anomaly~\cite{NuTeV:anomal}. Furthermore 
the strange quark contribution to the W-production at the LHC is of the 
same order of magnitude as the non-strange contribution.
These analyses include the (anti)neutrino data by the CCFR and NuTeV
collaborations in a global NLO fit to inclusive charged lepton-nucleon DIS and 
Drell-Yan data.
A strange sea suppression factor 
$\kappa = \int_0^1x[s(x)+\bar{s}(x)]{\rm d}x/\int_0^1x[\bar{u}(x)+\bar{d}(x)]{\rm d}x$ of $0.62\pm0.04({\rm exp})\pm0.03({\rm QCD})$ at $Q^2 = 20\,{\rm GeV}^2$ is obtained and the strange see is found to be 
slightly softer than the non-strange sea.
The  $x(s-\bar{s})$ asymmetry is compatible with zero within errors, 
as can bee seen in Fig.~\ref{Fig:ssbar}. 
The asymmetry changes sign when only including either CCFR or NuTeV data and is also 
sensitive to nuclear and higher order QCD corrections.

An extraction of  PDFs at high-$x$~\cite{PDF:hix}, a study performed in close collaboration with the 
CTEQ collaboration, was presented by Keppel. The goal of this study is to extend 
the standard CTEQ global PDF fit to larger values of $x$ and lower values of 
$Q^2$, taking advantage of a host of new results from Jefferson Lab experiments, 
as well as COMPASS and HERMES, which provide valuable high precision inputs in the high-$x$ 
region. Preliminary results of this study indicate that extracted PDFs 
are relatively insensitive to higher twist
 and target mass corrections as long as these corrections are simultaneously taken 
into account. The reduction of the PDF uncertainty at high $x$ in 
this preliminary study is quite encouraging. The indication 
of $d$-quark suppression at high $x$ is an intriguing suggestion.
Precise PDFs at high-$x$ are and will be an important input to current and upcoming 
neutrino oscillation experiments, and could also prove to be relevant for 
heavy particle production at the hadron colliders Tevatron and LHC.

The extraction of nuclear PDFs~\cite{Nucl:PDF}, presented by Paukkunen, 
aims for a process independent  nuclear PDF-set. The 
difference of free versus bound nucleons is taken into account 
by defining the bound proton PDFs in a nucleons of nucleon number $A$ 
relative to the free 
PDF as $f_i^A(x,Q^2) \equiv R_i^A(x,Q^2)f_i^{\rm CTEQ6.1M}( x,Q^2)$.
An excellent agreement between NLO pQCD and the hard process nuclear 
data for DIS, Drell-Yan and $\pi^0$ production over a wide kinematic 
range of $0.005<x<1$ and $1.69<Q^2<150\,{\rm GeV}^2$ is found, leading to the 
conclusion that the factorisation theorem in hard nuclear scattering processes 
works well. Further experiments, e.g. proton-ion collisions at the LHC or 
electron-ion colliders such as eRHIC and LHeC may be needed to find 
possible violations of the factorisation theorem.  

\section{Outlook on LHC}
The LHC start-up will bring particle physics to a new energy regime. 
The ATLAS, CMS and LHCb experiments are eagerly awaiting first data.
The first challenge will be to ``rediscover'' the Standard Model
at the LHC energy scale. Measuring the production of the $W$ and $Z$ bosons
will be an important milestone, which will test the understanding
of the detector calibrations as well as our knowledge of the Standard Model.

The uncertainties of the PDFs constitute the main source of theory uncertainty.
Measurements of the rapidity distribution of the $W$-decay leptons and the 
$W$-asymmetry in the early data of 100\,pb$^{-1}$ have the potential to improve 
our knowledge of PDFs, if the experimental systematics can be controlled with 
a precision better than 5\%. With increasing integrated luminosity a fit to 
the $Z$ and $W$ rapidity distributions should allow to measure the luminosity 
in the LHCb interaction region with a precision of one to two percent. 
A study of Drell-Yan events shows that LHCb has the capability to trigger and 
reconstruct Drell-Yan muon-pairs at a mass $M(\mu\mu) > 2.5\,{\rm GeV}^2$.
These events probe $x$-values down to $1.5\cdot10^{-6}$, i.e. $x$ values 
probed at HERA only at low values and limited range in $Q^2$. Therefore
the measurement could provide an important test of the extrapolation of PDFs
to much higher scales.

For further details of the LHC related studies presented by Anderson, Cepeda 
and De Lorenzi, see their contributions in these proceedings. 

\section{Theoretical Developments}

The theoretical talks in this working group can be naturally divided into two 
categories: those describing general aspects of structure functions and those 
focused on phenomena at low values of Bjorken $x$. In the former there were 
five contributions and in the latter eleven. We make a short description of 
each of them in the following two subsections.

\subsection{Structure Functions}

There were three contributions devoted to the investigation of the higher-loop 
structure of Feynman diagrams. 

Jos Vermaseren explained in detail what are 
the main obstacles when calculating structure functions at higher orders in 
the strong coupling. The natural space for these complicated calculations 
is the Mellin representation, this brings harmonic sums into the game. The 
inversion of these sum has as a consequence the appearance of harmonic 
polylogarithms. Many times these new functions arise after the application 
of a Mellin-Barnes technique which has the advantage of making the Lorenz 
integrations rather straightforward. Relations among the different harmonic 
polylogarithms allow for simplification of initially very complicated 
expressions. Of particular importance are the Multiple Zeta Values (MZV's) 
which are 
related to infinite sums of harmonic sums. To reduce them into a simple 
basis is a formidable task which has been at the hands of mathematicians 
from the 90's, useful tricks are the so-called 'shuffle' and 'stuffle' 
algebras. An important tool provided by Vermaseren and collaborators 
for the reduction of expressions is the code named as TFORM. He showed 
tables for the MZV's which cover all cases up to 6-loop coefficient 
functions. In Ref.~\cite{Blumlein:2009cf} further details can be found. 

Advanced reduction techniques using computers were also shown by Johannes 
Bl{\"u}mlein for the calculation of 3-loop anomalous dimensions and 
Wilson coefficients. He explained that single scale quantities depending on 
a ratio of Lorentz invariants can be transformed into a Mellin 
representation, when the Mellin-conjugate variable of this ratio is considered 
as discrete we can describe the calculation in terms of difference equations. 
In this way one can reconstruct the general formula up to 3-loop order 
for single scale quantities out of a finite number of fixed moments. 
This task is facilitated by the recurrence relations polynomials and 
nested harmonic sums fulfil. To solve the difference equation order by 
order they use the package ``sigma''. The main reference is Ref.~\cite{Blumlein:2009tj}.  

New higher-order results for 3-loop coefficient function of the structure 
function $F_3^{\nu+\bar{\nu}}$ were also discussed by Andreas Vogt. He 
remarked the appearance of new colour structures proportional to the 
$d^{abc} d_{abc}$ group invariant which are suppressed at $x \to 1$ but 
have a large effect at small $x$. For the coefficient functions for 
$F_{L}$ he remarked the very slow convergence of the perturbative 
expansion. See Ref.~\cite{Moch:2008fj,Moch:2009mu} for more details.  

A global analysis of parton distribution functions including a $p_T$ 
resummation was discussed by Hung-Liang Lai~\cite{Nadolsky:2008zw}. 
In this presentation the 
importance of the high precision measurement of the $W$ mass, as well 
as its $Q_T$ (due to the recoil of the $W$ in the transverse plane), was 
emphasised. A transverse momentum resummation of the leading logarithms 
generating the latter quantity was performed using the Collins-Soper-Sterman 
formalism. This also included some treatment of the non-perturbative higher 
twist effects. In the combined PDF and $p_T$ global analysis as new
inputs they also included the $p_T$ of Drell-Yan pairs and $Z$ bosons. 

A description of $F_L$ within the context of a colour-dipole model was 
provided by Dieter Schildknecht. He made an introduction to DIS at low $x$ 
from the point of view of colour-dipole interactions remarking the difference 
in the transverse momentum of $q {\bar q}$ pairs when produced from 
longitudinal or transverse virtual photons. As a consequence, the 
interesting relation $F_L = 0.27 F_2$ was derived with a minimal set of 
assumptions in their model. A description of geometric scaling was also 
provided. 

\subsection{Low-{\boldmath$x$}}

There were many talks and hot discussions in the low $x$ sessions. Felipe 
Llanes-Estrada described what he considered a loophole in the Deeply Virtual 
Compton Scattering (DVCS) factorisation theorems. For this they assumed, in 
the quark-nucleon scattering amplitude, a Regge behaviour independent of the 
number of quarks. This leads to a similar Regge form in structure functions 
and leads to a breakdown of the collinear factorisation in DVCS. 
The interested reader can consult Ref.~\cite{Brodsky:2009bp}. 

The difficult problem of a consistent treatment of unitarity corrections 
was addressed by Gian Paolo Vacca. If one wants to attack this problem 
using reggeized gluons as the correct degrees of freedom then it is very 
complicated since these are non local composite states. One can use a 
simplified effective field theory introduced by Gribov in the times before 
QCD. In this case the problem is reduced to zero transverse dimensions 
and is equivalent to a quantum mechanics of locally interacting pomerons. 
The model proposed by Vacca allows for the inclusion of pomeron loops 
induced by interactions. The relevant references are~\cite{Braun:2006gy,Vacca:2009tv}.

Many talks of phenomenological nature were focused on the application of 
evolution equations including DGLAP and BFKL type of resummations to 
the description of $F_{2,L}$. This is the case of the contribution of 
Anna Sta{\'s}to where she considers a DGLAP/BFKL unified approach to 
make predictions for $F_L$. The BFKL kernel is supplemented by a 
kinematic constraint, DGLAP splitting functions are included and 
running coupling effects taken into account. In this way both the 
collinear and dipole model limits are correctly obtained. Good agreement with 
HERA data was found and predictions for the LHeC were given, as can be found 
in Ref.~\cite{GolecBiernat:2009be}. 

In a similar spirit Henri Kowalski explained an interesting calculation 
of the gluon density at small $x$ using the BFKL equation modified in the 
infrared and improved by collinear effects. Their matching between the 
perturbative and non-perturbative regions discretizes the eigenfunctions 
of the kernel when the running of the coupling is considered, introducing 
phases which carry non-perturbative information. Playing with their values 
they managed to get accurate phenomenology for HERA data, see Ref.~\cite{Ellis:2008yp}.  

The renormalisation group improved BFKL equation underlies many of the 
different approaches presented in our working group. A very detailed 
description of it was given by Dimitri Colferai~\cite{Ciafaloni:2007gf}. 
He went into the 
technicalities of consistently including quarks into the game using a 
matrix form within a collinear factorisation scheme very close to 
$\overline{\rm MS}$ since the anomalous dimensions in this scheme are 
incorporated up to NLO. In the gluon channel the NLL BFKL kernel is 
included. Results for the resummed splitting function matrix were shown.

An alternative to the RG-improved kernels just summarised is that of 
exploiting the existing duality between the DGLAP and the BFKL approaches 
in the fixed coupling limit. The phenomenological implications of this 
idea were highlighted by Juan Rojo~\cite{Rojo:2009us}. The resummed $P_{gg}$ is very similar 
to that presented by Colferai and a smooth large $Q^2$ limit for 
$K$ factors is achieved. Results for $F_L$ and predictions for LHeC were 
also explained. 
   
The application of high energy resummations in Drell-Yan production was 
discussed by Simone Marzani. He explained how this cross section is one of 
the better known at the LHC and how it can be complemented by a small $x$ 
resummation in the large rapidity region. The calculation now 
requires the 
sum to the fixed order coefficient function of a tower of small $x$ logarithms 
and of subtraction terms to avoid double counting. All of this is performed 
in terms of Mellin moments.  At NLO the small $x$ corrections are of the 
order of a 5 to 10 percent, see Ref.~\cite{Marzani:2008uh,Marzani:2009hu}. 
These effects will be more sizable in rapidity distributions. 

As part of a joint session together with the Diffraction and Vector Mesons 
working group, Javier Albacete presented a global analysis of DIS inclusive 
structure function data using the BK 
equation~\cite{Albacete:2007yr,Albacete:2007sm,Albacete:2009fh}. 
They used a form of running 
the coupling based on a ratio of coupling functions at different scales 
related to the dipole size. As usual, the effect of the running is to reduce 
the onset of small $x$ corrections making the predictions more compatible with 
the experimental data. They also obtained a good fit of $F_L$ and made 
predictions for the LHeC.

An interesting talk by Emil Avsar was based on how to introduce an absorptive 
barrier in order to impose unitarity constraints in the CCFM equation, whose 
physical principle for resummation is angular ordering. This opens up the 
possibility to study unitarity effects in exclusive observable, an important 
road since it is very difficult to discriminate among different evolution 
approaches using only inclusive distributions. His approach is very flexible 
since it is already implemented in a Monte Carlo. 
Ref.~\cite{Avsar:2009pv,Avsar:2009pf} contains all the details. 

Finally, it was a great pleasure to attend the two more formal contributions 
of our working group related to the structure of the NLO corrections to the 
BK and BFKL equations, presented by Victor Fadin and Ian Balitsky. Fadin 
over-viewed the ambiguities present at NLO in the BFKL calculation. To be aware 
of the freedom in the definition of the formalism allows for a correct 
investigation of its conformal properties in the two dimensional transverse 
space. The game in the last few years has been to compare the original 
BFKL approach, based on the gluon Reggeization and bootstrap consistency 
conditions, with the colour dipole picture~\cite{Fadin:2009za}. The mapping between both is simple 
at LO but more complicated at NLO, however, taking into account the 
definition of the impact factors and the choice of the energy scale present 
in the logarithms of energy, it is possible to show that both approaches give 
equivalent evolution equations. 

The description of high energy scattering in terms of Wilson lines discussed 
by Balitsky is always very appealing. This allows for a direct calculation 
of evolution kernels in terms of colour dipoles propagating in the background 
of colour shock waves. Here the nonlinearity of the dipole evolution is 
immediately manifest. Recently the NLO corrections to the BK equation 
have been obtained, and can be found in Ref.~\cite{Balitsky:2008zza}. 

\section*{Acknowledgement}
We very much enjoyed this well organised workshop in the beautiful town of 
Madrid, and we thank the local organisers for their support and hospitality 
extended to us. We would also like to thank all those who contributed to the structure function session either by preparing talks or by taking part in the lively discussions.

%
%
%
%
%
%
%

\begin{footnotesize}




\begin{thebibliography}{99}
\bibitem{url} Slides: \\ 
\verb$http://indico.cern.ch/contributionDisplay.py?contribId=326&sessionId=22&confId=53294$\\
\verb$http://indico.cern.ch/contributionDisplay.py?contribId=15&sessionId=22&confId=53294$
\bibitem{H1:medQ2new} H1 Collaboration, F.D.~Aaron {\it et al.}, submitted to Eur.~Phys.~J.~C, arXiv:0904.3513 [hep-ex].
\bibitem{H1:lowQ2new} H1 Collaboration, F.D.~Aaron {\it et al.}, submitted to Eur.~Phys.~J.~C, arXiv:0904.0929 [hep-ex]. 
\bibitem{H1:lowQ2} H1 Collaboration, C.~Adloff {\it et al.}, Eur. Phys. J. {\bf C21} 33 (2001) [arXiv:0012053 [hep-ex]],\\
H1 Collaboration, C. Adloff {\it et al.} , Nucl. Phys. B 497, 3 (1997) [arXiv:9703012 [hep-ex]]. 
\bibitem{Dipole:GBW} K.~Golec-Biernat and M.~W\"usthoff, Phys. Rev. {\bf D59}, 014017 (1999) [arXiv:9807513 [hep-ph]].
\bibitem{Dipole:IIM} E.~Iancu, K.~Itakura and S.~Munier, Phys. Lett. {\bf B590}, 199 (2004) [arXiv:0310338 [hep-ph]].
\bibitem{Fractal} T. La\v stovi\v cka, Eur. Phys. J. {\bf C24}, 529 (2002) [arXiv:0203260 [hep-ph]].
\bibitem{H1:FL} H1 Collaboration, F.D.~Aaron {\it et al.}, Phys. Lett. {\bf B665}, 139 (2008) [arXiv:0805.2809 [hep-ex]]. 
\bibitem{ZEUS:FL} ZEUS Collaboration, S.~Chekanov {it et al.}, submitted to Phys. Lett. B, arXiv:09041092 [hep-ex].
\bibitem{ZEUSJETS} ZEUS Collaboration, S. Chekanov {\it et al.}, Eur. Phys. J. {bf C42}, 1 (2005) [arXiv:0503274 [hep-ph]].
\bibitem{H1PDF2000} H1 Collaboration, C. Adloff {\it et al.}, Eur. Phys. J. {\bf C30}, 1 (2003) [arXiv:0304003 [hep-ex]].
\bibitem{ZEUS:e-p:NC} ZEUS Collaboration, S.~Chekanov {it et al.}, accepted by Eur.~Phys.~J.~C, arXiv:09012385 [hep-ex].
\bibitem{ZEUS:e-p:CC} ZEUS Collaboration, S.~Chekanov {it et al.}, Eur.~Phys.~J. {\bf C61} 223 (2009) [arXiv:08124620 [hep-ex]].
\bibitem{H1:ZEUS:AVG} J. Feltesse, Proceedings of DIS 2008, doi:10.3360/dis.2008.24\\
A. Glazov, AIP Conf. Proc. {\bf 792}, 237 (2005)
\bibitem{ZEUS:BPC} ZEUS Collaboration, J. Breitweg {\it et al.}, Phys. Lett. {\bf B407}, 432 (1997),[arXiv:9707025 [hep-ex]]. 
\bibitem{ZEUS:BPT} ZEUS Collaboration, J. Breitweg {\it et al.}, Phys. Lett. {\bf B487}, 53 (2000) [arXiv:0005018 [hep-ex]].
\bibitem{ZEUS:SVX} ZEUS Collaboration, J. Breitweg {\it ey al.}, Eur. Phys. J. {\bf C7}, 609 (1999) [arXiv:9809005 [hep-ex]].
\bibitem{CDF:Jets:Kt} CDF Collaboration, A. Abulencia {\it et al.}, Phys. Rev. {\bf D75}, 092006 (2007),\\
Erratum ibid. {\bf D75}, 119901 (2007) [arXiv:0701051 [hep-ex]].
\bibitem{CDF:Jets:Mid} CDF Collaboration, A. Aaltonen {\it et al.}, Phys. Rev. {\bf D78}, 052006 (2008),\\
Erratum ibid. {\bf D79}, 119902 (2009) [arXiv:0807.2204 [hep-ex]].
\bibitem{D0:Jets} D\O\ Collaboration, V.M. Abazov {\it et al.}, Phys. Rev. Lett. {\bf 101}, 062001 (2008) [arXiv:0802.2400 [hep-ex]].
\bibitem{TeV:W:Tech} A. Bodek {\it et al.}, Phys. Rev. {\bf D77}, 111301 (2008) [arXiv:07112859 [hep-ph]].
\bibitem{HERAPDF01} H1 and ZEUS Collaborations, H1prelim-09-045, ZEUS-prel-09-011.\\
B. Reisert, Proceedings of the ICHEP 2008, arXiv:0809.4946 [hep-ex].
\bibitem{MSTW08} A.D. Martin {\it et al.}, Submitted to Eur. Phys. J. C, arXiv:0901.0002 [hep-ph].
\bibitem{CT09} J. Pumplin {\it et al.}, Phys. Rev. {\bf D80}, 014019 (2009), see also arXiv:0904.2424 [hep-ph] 
\bibitem{NNPDF} NNPDF Collaboration, R.D. Ball {\it et al.}, Nucl. Phys. {\bf B809}, 1 (2009), arXiv:0808.1231 [hep-ph], and references therein.
\bibitem{NNPDF:method} M. Ubiali, Nucl. Phys. Proc. Suppl. {\bf 186}, 62 (2009) [arXiv:0809.3716 [hep-ph]].
\bibitem{Strange:Alekhin} S. Alekhin {\it et al.}, Phys. Lett. {\bf B675}, 433 (2009) [arXiv:0812.4448 [hep-ph]].
\bibitem{Strange:Rojo} NNPDF Collaboration, R.D. Ball {\it et al.}, arXiv:09061958 [hep-ph].
\bibitem{NuTeV:anomal} S. Davidson {\it et al.}, JHEP {\bf 0202}, 037 (2002) [arXiv:0112302 [hep-ph]].
\bibitem{PDF:hix} J.F. Owens {\it et al.}, Phys. Rev. {\bf D75}, 054030 [arXiv:0702159 [hep-ph]].
\bibitem{Nucl:PDF} K.J. Eskola {\it et al.}, JHEP {\bf 0904}, 065 (2009) [arXiv:09024154 [hep-ph]].


\bibitem{Blumlein:2009cf}
  J.~Bl{\"u}mlein, D.~J.~Broadhurst and J.~A.~M.~Vermaseren,
  arXiv:0907.2557 [math-ph].

\bibitem{Blumlein:2009tj}
  J.~Bl{\"u}mlein, M.~Kauers, S.~Klein and C.~Schneider,
  arXiv:0902.4091 [hep-ph].

\bibitem{Moch:2008fj}
  S.~Moch, J.~A.~M.~Vermaseren and A.~Vogt,
  arXiv:0812.4168 [hep-ph].
\bibitem{Moch:2009mu}
  S.~Moch and A.~Vogt,
  JHEP {\bf 0904} (2009) 081
  [arXiv:0902.2342 [hep-ph]].

\bibitem{Nadolsky:2008zw}
  P.~M.~Nadolsky {\it et al.},
  Phys.\ Rev.\  D {\bf 78} (2008) 013004
  [arXiv:0802.0007 [hep-ph]].

\bibitem{Brodsky:2009bp}
  S.~J.~Brodsky, F.~J.~Llanes-Estrada, J.~T.~Londergan and A.~P.~Szczepaniak,
  arXiv:0906.5515 [hep-ph].

\bibitem{Braun:2006gy}
  M.~A.~Braun and G.~P.~Vacca,
  Eur.\ Phys.\ J.\  C {\bf 50} (2007) 857
  [arXiv:hep-ph/0612162].

\bibitem{Vacca:2009tv}
  G.~P.~Vacca,
  arXiv:0907.2581 [hep-ph].

\bibitem{GolecBiernat:2009be}
  K.~Golec-Biernat and A.~M.~Stasto,
  Phys.\ Rev.\  D {\bf 80} (2009) 014006
  [arXiv:0905.1321 [hep-ph]].

\bibitem{Ellis:2008yp}
  J.~Ellis, H.~Kowalski and D.~A.~Ross,
  Phys.\ Lett.\  B {\bf 668} (2008) 51
  [arXiv:0803.0258 [hep-ph]].

\bibitem{Ciafaloni:2007gf}
  M.~Ciafaloni, D.~Colferai, G.~P.~Salam and A.~M.~Stasto,
  JHEP {\bf 0708} (2007) 046
  [arXiv:0707.1453 [hep-ph]].

\bibitem{Rojo:2009us}
  J.~Rojo, G.~Altarelli, R.~D.~Ball and S.~Forte,
  arXiv:0907.0443 [hep-ph].

\bibitem{Marzani:2008uh}
  S.~Marzani and R.~D.~Ball,
  Nucl.\ Phys.\  B {\bf 814} (2009) 246
  [arXiv:0812.3602 [hep-ph]].

\bibitem{Marzani:2009hu}
  S.~Marzani and R.~D.~Ball,
  arXiv:0906.4729 [hep-ph].

\bibitem{Albacete:2007yr}
  J.~L.~Albacete and Y.~V.~Kovchegov,
  Phys.\ Rev.\  D {\bf 75} (2007) 125021
  [arXiv:0704.0612 [hep-ph]].

\bibitem{Albacete:2007sm}
  J.~L.~Albacete,
  Phys.\ Rev.\ Lett.\  {\bf 99} (2007) 262301
  [arXiv:0707.2545 [hep-ph]].

\bibitem{Albacete:2009fh}
  J.~L.~Albacete, N.~Armesto, J.~G.~Milhano and C.~A.~Salgado,
  arXiv:0902.1112 [hep-ph].

\bibitem{Avsar:2009pv}
  E.~Avsar and E.~Iancu,
  Phys.\ Lett.\  B {\bf 673} (2009) 24
  [arXiv:0901.2873 [hep-ph]].

\bibitem{Avsar:2009pf}
  E.~Avsar and E.~Iancu,
  arXiv:0906.2683 [hep-ph].

\bibitem{Fadin:2009za}
  V.~S.~Fadin, R.~Fiore and A.~V.~Grabovsky,
  Nucl.\ Phys.\  B {\bf 820} (2009) 334
  [arXiv:0904.0702 [hep-ph]].

\bibitem{Balitsky:2008zza}
  I.~Balitsky and G.~A.~Chirilli,
  Phys.\ Rev.\  D {\bf 77} (2008) 014019
  [arXiv:0710.4330 [hep-ph]].





\end{thebibliography}
%

\end{footnotesize}


\end{document}